\documentclass[english,prl,twocolumn,showpacs,superscriptaddress,preprintnumbers,amsmath,amssymb,a4paper]{revtex4}
\usepackage[T1]{fontenc}
\usepackage[latin1]{inputenc}
\usepackage{graphicx}
\usepackage[top=0.8in,bottom=0.9in,,left=0.7in,right=0.7in]{geometry}

\makeatletter
\usepackage{dcolumn}
\usepackage{bm}
\usepackage{epsfig}

\newcommand{\ignore}[1]{}

\usepackage{babel}
\makeatother

\begin{document}

\title{Orbit- and Atom-Resolved Spin Textures of Intrinsic, Extrinsic and Hybridized Dirac Cone States}

\author{Lin Miao \footnotemark[2]}
\affiliation{Key Laboratory of Artificial Structures and Quantum Control (Ministry of Education),
Department of Physics and Astronomy, Shanghai Jiao Tong University, Shanghai 200240, China}

\author{Z. F. Wang \footnotemark[2]}
\affiliation{Department of Materials Science and Engineering,University of Utah, Salt Lake City, UT 84112, USA}

\author{Meng-Yu Yao}
\author{Fengfeng Zhu}
\affiliation{Key Laboratory of Artificial Structures and Quantum Control (Ministry of Education),
Department of Physics and Astronomy, Shanghai Jiao Tong University, Shanghai 200240, China}
\author{J. H. Dil}
\affiliation{Swiss Light Source, Paul Scherrer Institute, CH-5232 Villigen, Switzerland}
\affiliation{Physik-Institut, Universit\"{a}t Z\"{u}rich-Irchel, 8057 Z\"{u}rich, Switzerland}
\author{C. L. Gao}
\author{Canhua Liu}
\affiliation{Key Laboratory of Artificial Structures and Quantum Control (Ministry of Education),
Department of Physics and Astronomy, Shanghai Jiao Tong University, Shanghai 200240, China}

\author{Feng Liu \footnotemark[1]}
\affiliation{Department of Materials Science and Engineering,University of Utah, Salt Lake City, UT 84112, USA}

\author{Dong Qian \footnotemark[1]}
\author{Jin-Feng Jia \footnotemark[1]}
\affiliation{Key Laboratory of Artificial Structures and Quantum Control (Ministry of Education),
Department of Physics and Astronomy, Shanghai Jiao Tong University, Shanghai 200240, China}


\begin{abstract}
Combining first-principles calculations and spin- and angle-resolved photoemission spectroscopy measurements, we identify the helical spin textures for three different Dirac cone states in the interfaced systems of a 2D topological insulator (TI) of Bi(111) bilayer and a 3D TI Bi$_2$Se$_3$ or Bi$_2$Te$_3$. The spin texture is found to be the same for the intrinsic Dirac cone of Bi$_2$Se$_3$ or Bi$_2$Te$_3$ surface state, the extrinsic Dirac cone of Bi bilayer state induced by Rashba effect, and the hybridized Dirac cone between the former two states.
Further orbit- and atom-resolved analysis shows that $s$ and $p_z$ orbits have a clockwise (counterclockwise) spin
rotation tangent to the iso-energy contour of upper (lower) Dirac cone, while $p_x$ and $p_y$ orbits have an
additional radial spin component. The Dirac cone states may reside on different atomic layers, but have the same spin texture.
Our results suggest that the unique spin texture of Dirac cone states is a signature
property of spin-orbit coupling, independent of topology.
\end{abstract}

\maketitle

\renewcommand{\thefootnote}{\fnsymbol{footnote}}
\footnotetext[2]{L.M. and Z.F.W. contributed equally to this work.}
\footnotetext[1]{Corresponding authors: fliu@eng.utah.edu; dqian@sjtu.edu.cn; jfjia@sjtu.edu.cn.}

One outstanding property that 3D TIs possess is the helical Dirac cone surface states residing inside
a bulk gap, in which electron spin is lock-in with momentum\cite{1,2}. This spin-momentum locking property
makes TIs promising materials for realizing spintronic devices, because the electron backscattering is
forbidden for nonmagnetic impurities. Therefore, understanding the spin texture of helical Dirac cone states
is of fundamental interest with practical implication.

Experimentally, the helical spin texture of surface Dirac cone state of 3D TIs has been directly detected by
spin- and angle-resolved photoemission spectroscopy (SARPES) \cite{3,4,5,6,7,8}. It can also be manipulated
through its interaction with different polarized lights \cite{prl,nature}. Very recently, it has been noted
that because of the strong spin-orbit coupling (SOC), total angular momentum, spin plus orbital angular momentum,
should be the good quantum number in TIs. Consequently, the spin texture of TIs is expected to be coupled with
atomic orbits in a specific manner \cite{9,10,XJZhou}. Both theory \cite{9} and experiment \cite{10,XJZhou} have shown some
very interesting patterns of radial/tangential spin texture coupled with $p$ orbits in Bi$_2$Se$_3$ family.

Besides the "intrinsic" Dirac cone state of TIs, "extrinsic" Dirac cone state may be generated by Rashba effect,
such as the one formed when Bi(111) bilayer is grown on Bi$_2$Se$_3$ substrate \cite{11,12}. Also, when Bi(111)
bilayer is grown on Bi$_2$Te$_3$ \cite{12}, a kind of hybridized Dirac cone state is formed between the intrinsic
surface Dirac cone state of Bi$_2$Te$_3$ and extrinsic Rashba Dirac cone state of Bi(111). These Dirac cone states
are all associated with strong SOC. However, different from the intrinsic surface states of TI, which are topologically
protected inside a bulk gap, the extrinsic and hybridized states have no topological origin, even though they are formed by interfacing a 2D TI of Bi(111) bilayer\cite{13,14,15,16} and a 3D TI of Bi$_2$Se$_3$ or Bi$_2$Te$_3$. It will be interesting to find out what spin texture such nontopological Dirac states have, in comparison with the topological surface Dirac states of 3D TIs.

In this Letter, the spin texture of three different Dirac cone states in the system of Bi/Bi$_2$Se$_3$ and
Bi/Bi$_2$Te$_3$ are systemically studied. Both first-principles calculations and SARPES measurements show that the
total spins form an identical helical spin texture for all three Dirac cones. Furthermore, the orbit-resolved calculations
reveal that $s$ and $p_z$ orbits have a clockwise (counterclockwise) spin rotation tangent to the iso-energy contour of
upper (lower) Dirac cone, while $p_x$ and $p_y$ orbits have a radial spin component, in agreement with recent studies \cite{9,10}. The atom-resolved calculations reveal that the Dirac cone states may reside on different atomic
layers, but their spin texture remains the same.

\begin{figure}
\includegraphics[width=8.5cm]{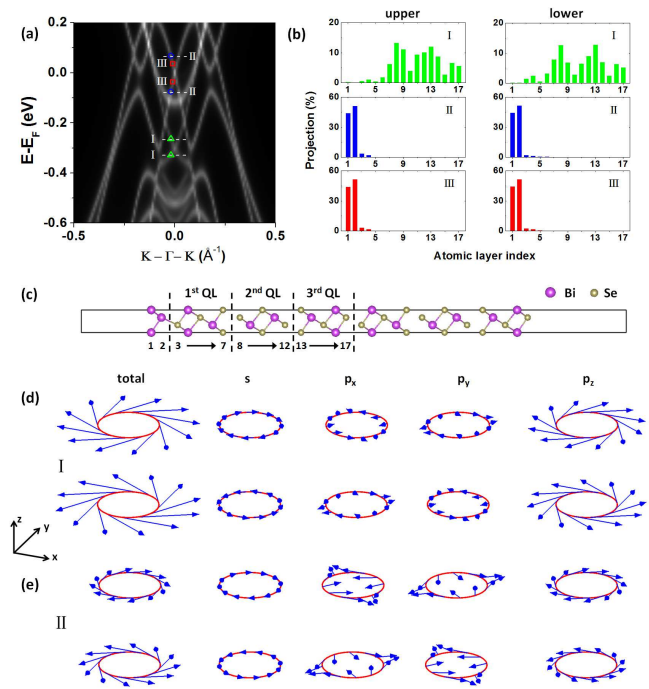}
\caption{(a) Bi/Bi$_2$Se$_3$ spectral function projected onto top Bi bilayer plus upper 2QL Bi$_2$Se$_3$.
(b) Percentage contribution of different atoms to the Dirac cone states marked by label I, II and III with
different colors and symbols in (a). Upper (lower) denotes upper (lower) Dirac cone states and the atomic
layer index is shown in (c). (c) Atomic structure of the optimized Bi/Bi$_2$Se$_3$. Bi bilayer
and upper 3QL Bi$_2$Se$_3$ are labeled with numbers to denote their atomic layer index. (d) and (e)
Orbit-resolved spin textures of the iso-energy contour Dirac cone state at the energy marked by label
I and II in (a).}
\end{figure}

Our first-principles calculations are carried out in the framework of generalized gradient
approximation with Perdew-Burke-Ernzerhof functional using the VASP package \cite{17}. The lattice
constants of the substrate are taken from experiments (a=4.138 {\AA} for Bi$_2$Se$_3$ and
a=4.386 {\AA} for  Bi$_2$Te$_3$), and Bi(111) bilayer is strained to match the substrate lattice
parameter. All calculations are performed with a plane-wave cutoff of 400 eV on an $11\times11\times1$
Monkhorst-pack k-point mesh. The substrate is modeled by a slab of 6 quintuple layer (QL) Bi$_2$Se$_3$
and Bi$_2$Te$_3$, and the vacuum layers are over 20 {\AA} thick to ensure decoupling between
neighboring slabs. During structural relaxation, atoms in the lower 4QL substrate are fixed in their
respective bulk positions, and Bi bilayer and upper 2QL of substrate are allowed to relax until
the forces are smaller than 0.01 eV/{\AA}.

First, we take a look at the Bi/Bi$_2$Se$_3$, where both intrinsic and extrinsic Dirac cone states coexist\cite{11}.
Figure 1(a) shows the spectral function of Bi/Bi$_2$Se$_3$ projected onto Bi bilayer plus top 2QL of
Bi$_2$Se$_3$. One sees two Dirac cones at the $\Gamma$ point, one below and the other near the Fermi level.
To reveal the origin of these two Dirac cones, we plot their real-space distribution by choosing three
data points from upper and lower Dirac cones, as marked by I, II and III in Fig. 1(a). The corresponding
momenta are -0.02, -0.02 and -0.01 {\AA}$^{-1}$ for I, II and III states, respectively. The atom-projected
state components are shown in Fig. 1(b), with the atomic layer indexes labeled in Fig. 1(c).
The Dirac cone states below the Fermi level come almost completely from the substrate with little contribution
from Bi bilayer (Fig. 1(b), top panel). This indicates that it is an intrinsic Dirac cone of Bi$_2$Se$_3$.
The Dirac cone states near the Fermi level come predominantly from the top Bi bilayer, having little substrate
contribution (see Fig. 1(b), two bottom panels). This indicates that it is an extrinsic Dirac cone generated
by the Rashba splitting of Bi bilayer bands, as discussed before \cite{11,15}. Additionally, we note that due
to interfacial interaction with the Bi bilayer, the intrinsic surface Dirac cone has a wide real-space distribution
in the 2nd and 3rd QL (Fig. 1(b), top panel) that peaks at the boundary between the 1st and 2nd, and the 2nd and 3rd QL (i.e.,
the 7th and 8th, and 12th and 13th atomic layer as seen in Fig. 1(c)). This is different from the surface Dirac cone of bare Bi$_2$Se$_3$
that decays from the 1st and 2nd QL \cite{18}. The Bi bilayer pushes down and broadens the surface states of
Bi$_2$Se$_3$. This is consistent with the recent findings by Wu \textit{et al.} that the vertical location of
surface Dirac cone state of a 3D TI can be tuned by depositing a layer of conventional insulator \cite{19}.

\begin{figure}
\includegraphics[width=8.5cm]{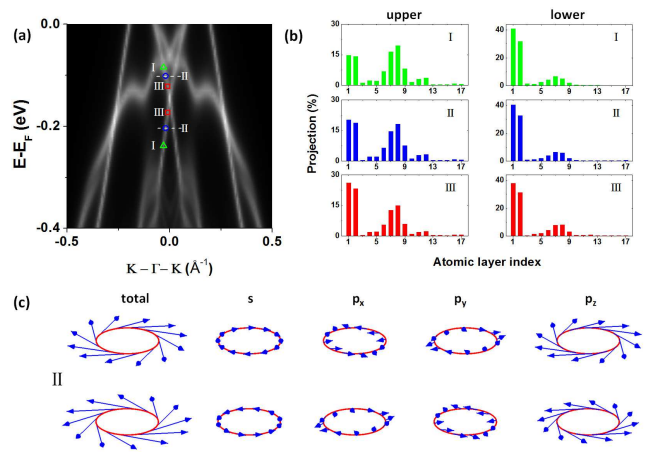}
\caption{(a) Bi/Bi$_2$Te$_3$ spectral function projected onto top Bi bilayer plus upper 1QL Bi$_2$Te$_3$.
(b) Percentage contribution of different atoms to the Dirac cone states marked by label I, II and III with
different colors and symbols in (a). (c) Orbit-resolved spin textures of the iso-energy contour Dirac cone state
at the energy marked by label II in (a). The scaling factor of the arrow length is the same to that in
Fig. 1.}
\end{figure}

\begin{figure*}
\includegraphics[width=17cm]{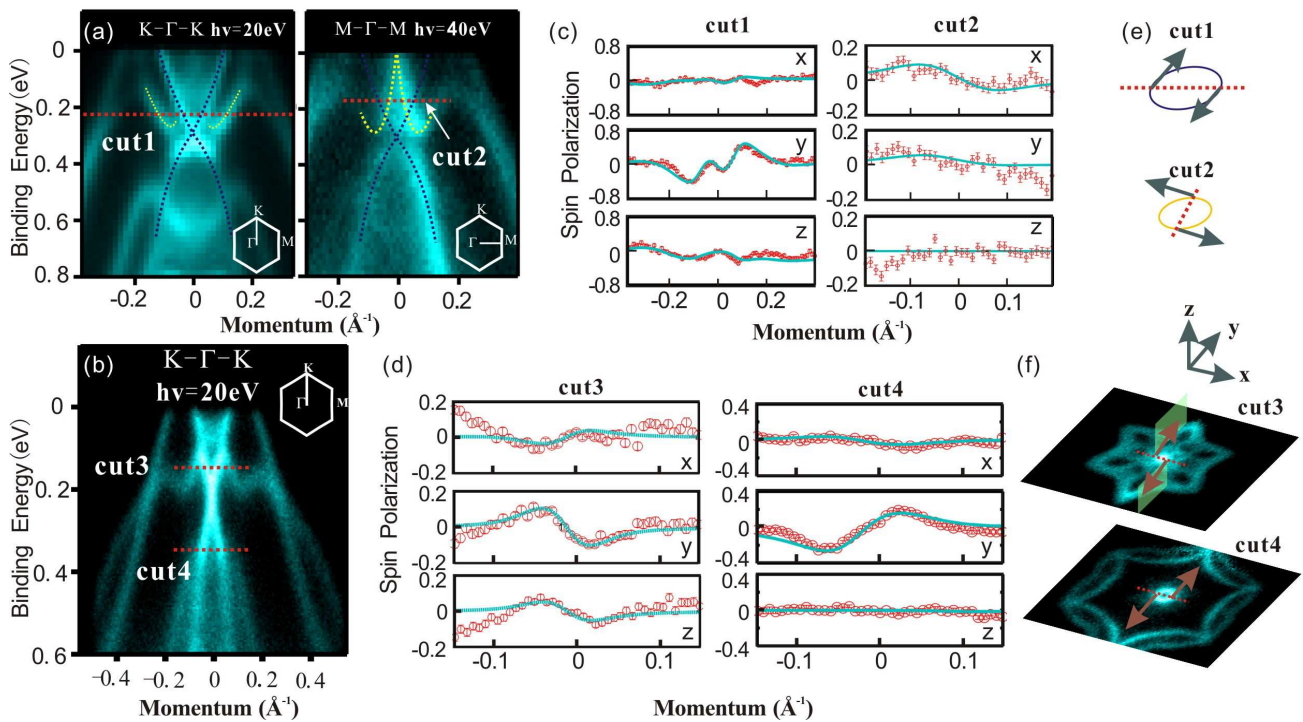}
\caption{(a)Spin-integrated ARPES spectra of Bi/Bi$_2$Se$_3$ and (b) Bi/Bi$_2$Te$_3$. The dashed blue and yellow lines show the position of the intrinsic and extrinsic Dirac cones. Experimentally, the relative intensity of two Dirac cones can be tuned by incident photon energy. (c) and (d) The experimental momentum dependent spin polarization and fitting data of Bi/Bi$_2$Se$_3$ and Bi/Bi$_2$Te$_3$. (e) and (f) Spin textures of intrinsic, extrinsic Dirac cones and hybridized Dirac cone. The images in (f) are the constant energy contours of ARPES intensity. }
\end{figure*}

Although the two Dirac cones in Fig. 1(a) have different physical origins, their spin textures are found
to be the same. The total and orbit-resolved spin textures along the iso-energy contour at the energy marked
by I (intrinsic Bi$_2$Se$_3$ Dirac cone) and II (extrinsic Bi Dirac cone) are shown in Fig. 1(d) and (e),
respectively. The total spins of both Dirac cones are oriented clockwise (counterclockwise) in upper (lower)
Dirac cone. $s$ and $p_z$ orbits have the same helical spin texture as the total spin, but $p_x$ and $p_y$ orbits
have a different spin texture with additional radial spin component, pointing toward opposite directions
along the 135$^\circ$ or (45$^\circ$) axis. In addition, $p_x$ orbit spin texture in upper (lower) Dirac cone is
same as $p_y$ orbit spin texture in lower (upper) Dirac cone. These in-plane $p$ orbit spin textures are consistent
with the theoretical model prediction by Zhang \textit{et al.} \cite{9}. We note that although both $p_x$ and
$p_y$ orbit spins have radial components, their summation, $p_x + p_y$ orbit spins is helical becoming tangent
to iso-energy contour, albeit they are orientated counterclockwise in both upper and lower Dirac cones. The
total helical in-plane p orbit spins with the same orientation in the upper and lower Dirac cones have been
recently experimentally observed in the Bi$_2$Se$_3$ surface Dirac cones\cite{XJZhou}. The only difference between the spin textures of intrinsic and extrinsic Dirac cones is the spin magnitude. For the intrinsic Dirac cone, the amplitude of $s$, $p_x$ and $p_y$ orbit spins are negligible, and the
magnitude of $p_z$ orbit spin is the largest. While for the extrinsic Dirac cone, the amplitude of $s$ orbit spin
is negligible, and the magnitude of three $p$ orbit spins are comparable to each other.

Next, we turn to the Bi/Bi$_2$Te$_3$ where hybridized Dirac cone states arise\cite{11}. Figure 2(a) shows the spectral function of Bi/Bi$_2$Te$_3$
projected onto Bi bilayer plus upper 1QL Bi$_2$Te$_3$. There appears only one Dirac cone at the $\Gamma$
point below Fermi level. Similar to the case of Bi/Bi$_2$Se$_3$, three groups of data
points are chosen around this Dirac cone, as marked by I, II and III in Fig. 2(a).
The corresponding momenta are -0.03, -0.02 and -0.01 {\AA}$^{-1}$ for I, II and III states, respectively.
The atom-projected components of these states are shown in Fig. 2(b), and the atomic layer indexes are
the same as those shown in Fig. 1(c). One major difference from Bi/Bi$_2$Se$_3$ is that the only one Dirac state
in Bi/Bi$_2$Te$_3$ has sizable contributions from both Bi bilayer and Bi$_2$Te$_3$ substrate, as clearly seen from Fig.
2(b). This indicates that it is a hybridized Dirac state between the intrinsic surface Dirac state of Bi$_2$Te$_3$ and the extrinsic Rashba Dirac
state of Bi bilayer. A previous study has further shown another interesting point that such hybridization might enhance many-body interaction
within the Dirac states\cite{11}. Figure 2(b) also shows that the hybridization in upper Dirac cone is stronger than
that in lower Dirac cone. Moving away from the Dirac point (Form III to I), the components of the Bi
bilayer (Bi$_2$Te$_3$ substrate) decreases (increases) in upper Dirac cone, but they show little change in
lower Dirac cone. The substrate contribution to the Dirac cone is not localized in the 1st QL but between the 1st and 2nd QL (i.e., the 6th to 9th atomic layer as seen in Fig. 1(c)). This is because the surface state is pushed slightly away from
the Bi/Bi$_2$Te$_3$ interface compared to the bare Bi$_2$Te$_3$ surface state \cite{8}. After confirming the hybridized nature of the Dirac cone
state in Bi/Bi$_2$Te$_3$, its spin textures along the iso-energy contour at the energy marked by label II are
shown in Fig. 2(d). The total and orbit-resolved spin textures of hybridized Dirac cone are the same as those in
the intrinsic and extrinsic Dirac cones of the Bi/Bi$_2$Se$_3$ system. $s$ and $p_z$ orbits have a helical spin
texture, while $p_x$ and $p_y$ orbits have a non-helical radial spin component.

We can conclude from the above calculation results several common features for the three different
Dirac cone states. (1) The total spins have a helical spin texture, which is clockwise (counterclockwise) in upper (lower)
Dirac cone. (2) $s$ and $p_z$ orbit spins have the same texture as to the total spins. (3) $p_x$ and $p_y$ orbit spins
have a radial spin component individually, but their sum becomes helical. (4) The substrate surface state is
pushed down away from the 1st QL by the presence of an overlayer of Bi bilayer. 

To support our first-principles calculations, we have measured the total spin textures of these Dirac cone states in the Bi/Bi$_2$Se$_3$ and
Bi/Bi$_2$Te$_3$ systems by SARPES. The method for epitaxially growing high quality Bi/Bi$_2$Te$_3$ and Bi/Bi$_2$Se$_3$
samples are same to our previous works \cite{11,16}. SARPES measurements were performed at the SIS beam line at the Swiss Light Source using the
COPHEE spectrometer with a single 40 kV classical Mott detector. The typical energy and momentum resolution was 50 meV and 3\% of the surface Brillouin zone at SIS. All the measurements were carried out at 30K in ultrahigh vacuum with a base pressure better than $1\times10^{-10}$ torr. Proper measurement geometries were used to minimize SARPES matrix elements effects. For Bi/Bi$_2$Se$_3$ system, though there are two Dirac cone states, experimentally we tune the relative spectra intensity by using different photon energies. As shown in Fig. 3(a), the intrinsic Dirac cone from Bi$_2$Se$_3$ was observed clearly with photon energy of 20eV and the extrinsic Dirac Con from Bi bilayer was detected with photon energy of 40eV.

Figure 3(a) shows the spin-integrated ARPES spectra of Bi/Bi$_2$Se$_3$ along K-$\Gamma$-K and K-M-K directions, in which one
can see both intrinsic and extrinsic Dirac cones. In the SARPES measurements, two spin-resolved momentum distribution curves (MDC) were taken (cut1 and cut2, as shown by the horizontal dashed lines in Fig. 3(a)). Cut1 is at 220 meV below the Fermi level to study the intrinsic Dirac cone, and cut2 is at 180 meV below the Fermi level to study the extrinsic Dirac cone. The spin polarization is extracted by using the the two-step fitting routine of SARPES data\cite{3,fitting,20}. The momentum dependent spin polarization and fitting data are shown in Fig. 3(c), from which we extract the spin textures. Figure 3(e) presents the spin textures of intrinsic and extrinsic Dirac cones, which is clockwise in upper intrinsic Dirac cone and counterclockwise in lower extrinsic Dirac cone. For both Dirac cone, spins are nearly in-plane polarized. Figure 3(b) shows the spin-integrated ARPES spectra of Bi/Bi$_2$Te$_3$ along K-$\Gamma$-K direction, in which one can see the hybridized Dirac cone. Similar to the SARPES measurement for Bi/Bi$_2$Se$_3$, we also take two MDC cuts (cut3 and cut4), as shown by the dashed lines in Fig. 3(c). Cut3 (cut4) is taken at 150 (350) meV below the Fermi level, slightly above (below) the hybridized Dirac cone. The polarization and fitting data are shown in Fig. 3(d). As shown in Fig. 3(f), the spin texture is clockwise and counterclockwise in upper and lower Dirac cone, respectively. Thus, our SARPES measurements directly identify that all
three Dirac cones have the same helical spin textures, independent of topology, consistent with our first-principles
calculations.

Lastly, we present additional first-principles calculations to show the orbit- and atom-resolved spin textures of the
Dirac cone states. Here, we will focus on the top two Bi atoms. Overall, except the spin magnitude, the atom-resolved spin
textures have the same structure compared to the spin textures by adding all atoms together, as shown
in Fig. 4. However, from the atom projection, we can obtain some detailed local information of spin texture by
considering the spin magnitude difference. Figure 4(a) and (b) are the spin textures of extrinsic Dirac
cone projected onto the top two Bi atoms of Bi/Bi$_2$Se$_3$. We can see that the spin magnitude is very
different between the two Bi atoms. For the Bi-1 atom, $p_z$ orbit spin has the largest magnitude, while $s$,
$p_x$ and $p_y$ orbit spins are negligible. But for the Bi-2 atom, $p_x$ and $p_y$ orbit spins have the
largest magnitude, while $s$ and $p_z$ orbit spins are negligible. Figure 4(c) and (d) are the spin
textures of hybridized Dirac cone projected onto the top two Bi atoms of Bi/Bi$_2$Te$_3$. The atom-resolved
spin textures of hybridized Dirac cone are the same to those in the extrinsic Dirac cone. We
have also checked the atom-resolved spin textures for the intrinsic Dirac cone, which are again the same.

\begin{figure}
\includegraphics[width=8.5cm]{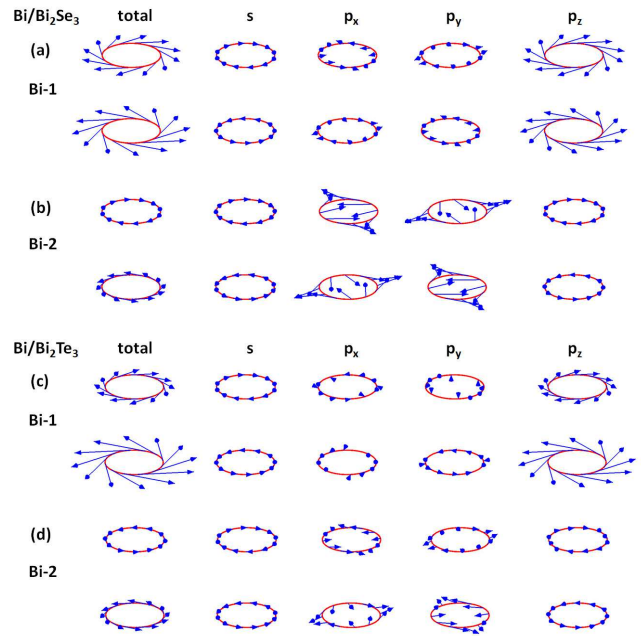}
\caption{Atom- and orbit-resolved spin textures of the iso-energy contour Dirac cone state at the
energy marked by label II in Fig. 1(a) and Fig. 2(a). (a) and (b) are the spin textures of
the first (Bi-1) and second (Bi-2) atom in Bi/Bi$_2$Se$_3$. (c) and (d) are the spin textures of
the first (Bi-1) and second (Bi-2) atom in Bi/Bi$_2$Te$_3$. The scaling factor of the arrow length
is twice of that in Fig. 1.}
\end{figure}

In summary, we have identified the helical spin textures for three different Dirac cone states in
the interfaced systems of Bi/Bi$_2$Se$_3$ and Bi/Bi$_2$Te$_3$, from first-principles calculations
and experiments. We confirm the recent theory \cite{9} and experiment \cite{10} that the spin texture of intrinsic surface Dirac cone
states of 3D TIs is coupled with orbit, resulting in an overall spin-orbit texture. Characteristically,
$s$ and $p_z$ orbits have the conventional helical spin texture; $p_x$ and $p_y$ orbits show individually
radial spin component, while the sum of the two shows a total in-plane helical spins. We further show that
the same spin texture is also applicable to the extrinsic Rashba Dirac cone states of Bi(111) bilayer on
a substrate as well as to the hybridized Dirac cone states between a TI surface state and a thin-film
Rashba state. Therefore, we suggest that the unique spin-orbit texture of helical Dirac states is
pertained to SOC, but not necessarily to TIs.

The experimental work conducted at Shanghai Jiao Tong University is supported by National
Basic Research Program of China Grants 2012CB927401, 2011CB921902, and 2011CB922200; National
Natural Science Foundation of China Grants 91021002, 10904090, 11174199, 11134008, and 11274228;
and Shanghai Committee of Science and Technology, China Grants 10QA1403300, 10JC1407100,
10PJ1405700, 12JC1405300 and 13QH1401500. The theoretical work conducted at University of Utah is supported by
the office of Basic Energy Sciences, US Department of Energy Grant DE-FG02-04ER46148. We also thank the
CHPC at University of Utah and NERSC for providing the computing resources. D.Q. acknowledges additional
support from Program for Professor of Special Appointment (Eastern Scholar)
at Shanghai Institutions of Higher Learning. Z.F.W. acknowledges additional support from
NSF-MRSEC (Grant No. DMR-1121252) and ARL (Cooperative agreement No. W911NF-12-2-0023).


\begin{references}

\bibitem{1}
M. Z. Hasan, C. L. Kane,
Rev. Mod. Phys. \textbf{82}, 3045 (2010).

\bibitem{2}
X.-L. Qi, S.-C. Zhang,
Rev. Mod. Phys. \textbf{83}, 1057 (2011).

\bibitem{3}
D. Hsieh, Y. Xia, L. Wray, D. Qian, A. Pal, J. H. Dil, J. Osterwalder, F. Meier,
G. Bihlmayer, C. L. Kane, Y. S. Hor, R. J. Cava, and M. Z. Hasan,
Science \textbf{323}, 919 (2009).

\bibitem{4}
D. Hsieh, Y. Xia, D. Qian, L. Wray, J. H. Dil, F. Meier, J. Osterwalder, L. Patthey, J. G. Checkelsky,
N. P. Ong, A. V. Fedorov, H. Lin, A. Bansil, D. Grauer, Y. S. Hor, R. J. Cava, and M. Z. Hasan,
Nature \textbf{460}, 1101 (2009).

\bibitem{5}
P. Roushan, J. Seo, C. V. Parker, Y. S. Hor, D. Hsieh, D. Qian,
A. Richardella, M. Z. Hasan, R. J. Cava, and A. Yazdani,
Nature \textbf{460}, 1106 (2009).

\bibitem{6}
Z.-H. Pan, E. Vescovo, A. V. Fedorov, D. Gardner, Y. S. Lee, S. Chu, G. D. Gu, and T. Valla,
Phys. Rev. Lett. \textbf{107}, 257004 (2011).

\bibitem{7}
S. Souma, K. Kosaka, T. Sato, M. Komatsu, A. Takayama, T. Takahashi, M. Kriener, K. Segawa, and Y. Ando,
Phys. Rev. Lett. \textbf{106}, 216803 (2011).

\bibitem{8}
S. V. Eremeev, G. Landolt, T. V. Menshchikova, B. Slomski, Y. M. Koroteev, Z. S. Aliev, M. B. Babanly, J. Henk,	
A. Ernst, L. Patthey, A. Eich, A. A. Khajetoorians, J. Hagemeister, O. Pietzsch, J. Wiebe, R. Wiesendanger,	
P. M. Echenique, S. S. Tsirkin, I. R. Amiraslanov, J. Hugo Dil, and Evgueni V. Chulkov,
Nature Commun. \textbf{3}, 635 (2012).

\bibitem{prl}
C.-H. Park, S. G. Louie,
Phys. Rev. Lett. \textbf{109}, 097601 (2012).

\bibitem{nature}
C. Jozwiak, C.-H. Park, K. Gotlieb, C. Hwang, D.-H. Lee, S. G. Louie, J. D. Denlinger, C. R. Rotundu, R. J. Birgeneau,	
Z. Hussain, and A. Lanzara, Nature Phys. \textbf{9}, 293 (2013).

\bibitem{9}
H. Zhang, C.-X. Liu, and S.-C. Zhang, Phys. Rev. Lett. \textbf{111}, 066801 (2013).

\bibitem{10}
Y. Cao, J. A. Waugh, X.-W. Zhang, J.-W. Luo, Q. Wang, T. J. Reber, S. K. Mo, Z. Xu, A. Yang, J. Schneeloch,	
G. D. Gu, M. Brahlek, N. Bansal, S. Oh, A. Zunger, and D. S. Dessau,
Nature Phys. \textbf{9}, 499 (2013).

\bibitem{XJZhou}
Zhoujin Xie et. al., arXiv:1303.0698 (2013).

\bibitem{11}
L. Miao, Z. F. Wang, W. Ming, M.-Y. Yao, M.-X. Wang, F. Yang, Y. R. Song, F. Zhu, A. V. Fedorov, Z. Sun, C. L. Gao,
C. Liu, Q.-K. Xue, C.-X. Liu, F. Liu, D. Qian, and J.-F. Jia,
Proc. Natl. Acad. Sci. USA \textbf{110}, 2758 (2013).

\bibitem{12}
Z. F. Wang, M.-Y. Yao, W. Ming, L. Miao, F. Zhu, C. Liu, C. L. Gao, D. Qian, J.-F. Jia, and F. Liu,
Nature Commun. \textbf{4}, 1384 (2013).

\bibitem{13}
S. Murakami, Phys. Rev. Lett. \textbf{97}, 236805 (2006).

\bibitem{14}
Z. Liu, C.-X. Liu, Y. S. Wu, W. H. Duan, F. Liu, and J. Wu,
Phys. Rev. Lett. \textbf{107}, 136805 (2010).

\bibitem{15}
L. Chen, Z. F. Wang, and F. Liu,
Phys. Rev. B \textbf{87}, 235420 (2013).

\bibitem{16}
F. Yang, L. Miao, Z. F. Wang, M.-Y. Yao, F. Zhu, Y. R. Song, M.-X. Wang, J.-P. Xu, A. V. Fedorov, Z. Sun,
G. B. Zhang, C. Liu, F. Liu, D. Qian, C. L. Gao, and J.-F. Jia,
Phys. Rev. Lett. \textbf{109}, 16801 (2012).

\bibitem{17}
G. Kresse, J. Hafner,
Phys. Rev. B \textbf{47}, 558 (1993).

\bibitem{18}
W. Zhang, R. Yu, H.-J. Zhang, X. Dai, and Z. Fang,
New J. Phys. \textbf{12}, 065013 (2010).

\bibitem{19}
G. Wu, H. Chen, Y. Sun, X. Li, P. Cui, C. Franchini, J. Wang, X.-Q. Chen, and Z. Zhang,
Sci. Rep. \textbf{3}, 1233 (2013).

\bibitem{fitting}
F. Meier et al., Phys. Rev. B 77, 165431 (2008).

\bibitem{20}
J. H. Dil,
J. Phys.: Condens. Matter \textbf{21}, 403001 (2009).


\end{references}
\end{document}